\documentclass{aa}

\usepackage{txfonts}
\usepackage{url}
\usepackage{hyperref}
\usepackage{orcidlink}
\usepackage{graphicx}

\newcommand{\dg}{^{\circ}}

\begin{document}

   \title{Polarization of active galactic nuclei with significant VLBI-\textit{Gaia} displacements\thanks{New polarization data discussed in this paper are only available in electronic form at the CDS via anonymous ftp to cdsarc.u-strasbg.fr (130.79.128.5) or via \url{http://cdsweb.u-strasbg.fr/cgi-bin/qcat?J/A+A/}.}}


   \author{Dmitry Blinov
          \inst{1,2}\orcidlink{0000-0003-0611-5784} 
          \and
          Arina Arshinova\inst{3,4}
          }

   \institute{Institute of Astrophysics, Foundation for Research and Technology-Hellas, GR-70013 Heraklion, Greece
   \and
   Department of Physics, University of Crete, GR-70013, Heraklion, Greece\\
              \email{blinov@ia.forth.gr}
         \and
             Special Astrophysical Observatory, Russian Academy of Sciences, Nizhny Arkhyz, 369167, Russia
        \and
             St. Petersburg University, 7/9 Universitetskaya nab., St.Petersburg, 199034, Russia\\
             }

   \date{Received June 20, 2024; accepted September 17, 2024}

 
  \abstract
   {Numerous studies have reported significant displacements in the coordinates of active galactic nuclei (AGNs) between measurements using the very-long-baseline-interferometry (VLBI) technique and those obtained by the \textit{Gaia} space observatory. There is consensus that these discrepancies do indeed manifest astrometrically resolved sub-components of AGNs rather than random measurement noise. Among other evidence, it has been reported that AGNs with VLBI-to-\textit{Gaia} displacements (VGDs) pointing downstream of their parsec-scale radio jets exhibit higher optical polarization compared to sources with the opposite (upstream) VGD orientation.}
   {We aim to verify the previously reported connection between optical polarization and a VGD-jet angle using a larger dataset of polarimetric measurements and updated \textit{Gaia} DR3 positions. We also seek further evidence supporting the disk-jet dichotomy as an explanation of such a connection by using millimeter-wave polarization and multiband optical polarization measurements.}
   {We performed optical polarimetric observations of 152 AGNs using three telescopes. These data are complemented by other publicly available polarimetric measurements of AGNs. We cross-matched public astrometric data from VLBI and \textit{Gaia} experiments, obtained corresponding positional displacements, and combined this catalog with the polarimetric and jet direction data.}
   {Active galactic nuclei with downstream VGDs are confirmed to have significantly higher optical fractional polarization than the upstream sample. At the same time, the millimeter-wavelength polarization of the two samples shows very similar distributions.}
   {Our results support the hypothesis that the VGDs pointing down the radio jet are likely caused by a component in the jet emitting highly polarized synchrotron radiation and dominating in the overall optical emission. The upstream-oriented VGDs are likely to be produced by the low-polarization emission of the central engine's subcomponents, which dominate in the optical.}


   \maketitle
%

\section{Introduction} \label{sec:intro}
Until recently, only a handful of active galactic nuclei (AGNs) with optically resolved jets, predominantly on the kiloparsec scale, were known \citep{Whiting2000}. Following the deployment of the European Space Agency's (ESA) \textit{Gaia} astrometry mission \citep{Gaia2016}, the landscape underwent a profound transformation. Presently, there is a potential to probe thousands of AGNs, discerning milliarcsecond (mas) substructures in the optical band.

Soon after the first \textit{Gaia} data release, it was discovered that many AGNs exhibit significant displacements between very-long-baseline-interferometry (VLBI) and \textit{Gaia} positions, with amplitudes reaching tens of milliarcseconds \citep{Petrov2017a, Petrov2017b, Makarov2019, Petrov2019, Klioner2022}. It has been demonstrated that when a radio jet is detected, these displacements tend to align with the jet's direction \citep{Kovalev2017, Xu2021}. This suggests that either the extended jet or the accretion disk dominates the overall optical AGN emission, causing the \textit{Gaia} photocenter to shift downstream or upstream along the jet with respect to the radio core. \cite{Plavin2019} demonstrated that the direction and amplitude of the VLBI-to-\textit{Gaia} displacement (VGD) vector are not only linked to the jet direction, but also show complex dependencies on the type, color index of the optical counterpart, redshift, and other AGN parameters. Based on these correlations, they concluded that a significant fraction of the VGD is physical in nature and is caused by the ability to resolve subcomponents in AGNs, made possible by \textit{Gaia}. \cite{Secrest2022} found that the prevalence of statistically significant optical-radio offsets is inversely correlated with photometric variability. Additionally, \cite{Lambert2024b} recently demonstrated that VLBI astrometric variability is also correlated with optical-radio offsets. Finally, there is evidence suggesting that the optical centroid often coincides with stationary or quasi-stationary features in the radio jet \citep{Lambert2024}.

Given that the direction of the VGD vector indicates whether the jet or the central engine (comprising the combined emission of the accretion disk, broad- and narrow-line regions, etc.) dominates in the overall optical emission of an AGN, one would naturally anticipate a statistical correlation between the VGD and polarization parameters. This correlation arises from the nature of the emission processes involved. The jet emits predominantly through the synchrotron process, leading to high optical polarization \citep{Ginzburg1979}. Conversely, the central engine emission is mostly thermal in nature. Its polarization is caused by scattering and is typically low \citep{Agol1997}. This expectation is particularly relevant for sources where the VGD can be accurately determined, which are typically Doppler boosted quasars observed in close proximity to their axis of symmetry. Such a dependence was indeed confirmed by \cite{Kovalev2020}, who found that AGNs with downstream-pointing VGDs have higher optical polarization compared to their upstream peers.

Understanding the displacement between VLBI and \textit{Gaia} positions is crucial for two key reasons. Firstly, it has the potential to probe the previously inaccessible substructure of AGNs at milliarcsecond scales in the optical band. This capability can help distinguish the disk and the jet emission, thereby offering solutions to various AGN physics problems. Secondly, it facilitates the implementation of more effective weighting strategies and enables the selection of optimal sources for aligning the celestial reference frames in both radio and optical domains.

This paper builds on the study by \cite{Kovalev2020}. Here, we utilize the \textit{Gaia} DR3 catalog \citep{GaiaDR3}, which offers higher precision compared to the previously used DR2. Additionally, we employed archival millimeter-band and new optical polarimetric data, as described in the subsequent section.

\section{Observations, data, and their reduction} \label{sec:data}

We conducted optical polarimetric observations of 152 AGNs using three telescopes and merged this catalog with archival data. The observing sample was selected from the list of AGNs with significant VGDs provided by \cite{Kovalev2020}, which is based on \textit{Gaia} DR2 positions. We applied visibility constraints and brightness thresholds specific to each telescope's limiting magnitude to the initial source list. Furthermore, we excluded sources with prominent host galaxies, as a large VGD in such cases could potentially result from non-axisymmetric brightness distributions within these galaxies \citep{Makarov2019}.


\subsection{FORS2 polarization data} \label{subsec:FORS}

We observed 47 AGNs utilizing FORS2 at the Very Large Telescope (VLT) of the European Southern Observatory (ESO). The observations took place between November 2019 and March 2020 (program id: 0104.B-0351). We used the R\_SPECIAL+76 filter and four half-wave plate (HWP) positions for each target. The instrumental polarization was monitored through measurements performed within the standard ESO calibration scheme. Using measurements of the standard stars WD0310$-$688, WD1344+106, WD1620$-$391, and WD2359$-$434 observed on dates adjacent to our observations, we did not find any systematic instrumental polarization offset exceeding 0.05\%. Since the photon noise uncertainties of our program targets are typically much larger, around 0.2\% in fractional polarization, we considered the instrumental polarization negligible and did not apply any correction for it.

For data processing, we developed a dedicated pipeline in Python leveraging the NumPy\footnote{\url{https://numpy.org/}}, AstroPy\footnote{\url{https://www.astropy.org/}}, AstroScrappy,\footnote{\url{https://github.com/astropy/astroscrappy}} and PhotUtils\footnote{\url{https://photutils.readthedocs.io/}} libraries. This pipeline executes standard preliminary processing steps including flat-fielding, de-biasing, cosmic-ray removal, and frame summation with sub-pixel accuracy (in cases where a long exposure was split into multiple shorter exposures). Ordinary and extraordinary ray measurements were conducted using the standard aperture photometry technique, with identical aperture sizes for both rays. The optimal aperture size was determined by maximizing the signal-to-noise ratio of the derived Stokes parameters while ensuring their stability under slight variations in aperture size around the optimal value. The polarization parameters were derived using equations from \cite{Patat2006}.

\subsection{RoboPol polarization data} \label{subsec:RoboPol}

We observed 92 AGNs between July 2017 and July 2022 with the RoboPol polarimeter at the 1.3-m telescope of the Skinakas Observatory\footnote{\url{https://skinakas.physics.uoc.gr/}} in Greece. Most of the sources were observed only once in the Cousins-R band, while a few were measured on multiple (up to 7) epochs. Additionally, 21 sources were also observed in the SDSS-$g^\prime$ band quasi-simultaneously with the R-band measurements. The polarimeter was developed specifically for observations of quasars in polarization \citep{Blinov2021a}. It consists of a combination of two HWP and two Wollaston prisms oriented in such a way that the source of interest is imaged on the detector by four rays with the polarization plane rotated by 45 degrees with respect to each other. The data were processed with the standard RoboPol pipeline \citep{King2014}, and instrumental polarization was corrected using standard stars observations as described in \cite{Blinov2023}.

\subsection{CAFOS polarization data} \label{subsec:CAHA}

We observed 18 AGNs in the Cousins-R and SDSS-$g^\prime$ bands using the 2.2-meter telescope at the Calar Alto Observatory (CAHA) in Spain, utilizing the multimode optical instrument Calar Alto Faint Object Spectrograph (CAFOS). The observations were performed on the nights of November 28 and 29 and December 30, 2019 (program id: OPTICON 2019B/029). As with the other instruments, the imaging polarimetry mode was used. Most of the program sources were measured with four positions of the rotating half-wave plate.

For the processing of the data, we adapted the same pipeline used for FORS2 data (see Sect.~\ref{subsec:FORS}). Among the program sources, a set of polarization standard stars including BD+28.4211, BD+59.389, G191B2B, and HD14069 was observed. By measuring the Stokes parameters of these stars and their deviation from the catalog values \citep{Blinov2023}, we determined the instrumental polarization and the polarization angle zero point. The uncertainties in the standards measurements were propagated to the final polarization parameters of the AGN.

\subsection{Archival polarization and jet orientation data} \label{subsec:arch}

Our observing data were complemented with archival optical polarization data collected and analyzed by \cite{Kovalev2020} and \cite{Friedman2020}. The former work is based on the data presented in \cite{Hutsemekers2005,Hutsemekers2018} and measurements obtained in two blazar monitoring programs \citep[Kanata, ][]{Itoh2016} and \citep[RoboPol, ][]{Blinov2021a}. This dataset has the advantage of being observed relatively close in time to the \textit{Gaia} observation period. Moreover, the polarization parameters for many sources in this set are calculated as averages among tens or hundreds of individual measurements distributed over years of observations. Therefore, these observations represent the average polarization of sample sources rather than a random moment snapshot. Since \textit{Gaia} DR3 measurements are based on observations spanning a period of 34 months, their collation with average polarization data is expected to be more accurate. On the other hand, \cite{Friedman2020} compiled the largest AGN catalog with optical broadband polarimetric measurements from 23 references in the literature. Despite being less uniform and dating back to 1983, this catalog offers the advantage of a larger sample size. Taking these differences into account, in Sect.~\ref{sec:pol_vs_vg}, we analyze two datasets independently. The first dataset includes our new measurements combined with data from \cite{Kovalev2020}. The second, referred to as the entire polarimetric sample, comprises our new measurements, data from \cite{Kovalev2020}, and data from \cite{Friedman2020}, with the merging order prioritized as listed.

We also utilized the millimeter-band polarization data from the \textit{Planck} Catalog of Polarized and Variable Compact Sources by \cite{Rocha2023}. These measurements were obtained with the \textit{Planck} space observatory of the European Space Agency in the frequency range of 30-353 GHz. We chose to use only the 44 GHz frequency channel data since it has the highest number of nonzero polarization measurements among the cataloged sources.

For the parsec-scale jet position angle (PA) data, we combined catalogs of AGN jet directions from \cite{Mandarakas2021} and \cite{Plavin2022}. Both of these works use the Astrogeo database \footnote{\url{https://astrogeo.org/}} data, but they employ different techniques to derive the jet PA. For sources present in both catalogs, we used the jet PA from the former paper. Overall, the two catalogs show good agreement, as demonstrated by \cite{Plavin2022}. The typical uncertainty of these PA measurements is $\sim 10 \dg$ \citep{Blinov2020}.
Redshifts were obtained from the NASA/IPAC Extragalactic Database\footnote{\url{https://ned.ipac.caltech.edu/}} and the Optical Characteristics of Astrometric Radio Sources \citep[OCARS,][]{Malkin2018} catalog.

\subsection{Astrometric data} \label{subsec:astrom}

We matched the radio positions of the following AGNs reported in publicly available catalogs from three astrometric/geodetic VLBI projects with their optical
counterparts presented in \textit{Gaia} DR3 \citep{GaiaDR3}: ICRF3 \citep{Charlot2020}, OPA2023a\footnote{\url{https://ivsopar.obspm.fr/}}, and RFC2022c.\footnote{\url{http://astrogeo.org/sol/rfc/rfc_2022c/}}   The ICRF3 provides three separate catalogs based on solutions obtained at different frequencies or their combinations. For sources presented in more than one of these catalogs, we used coordinates in the S/X (2.3/8.4 GHz), K (24 GHz), and X/Ka (8.4/32 GHz) bands, in that order, which is of decreasing priority.

The combined radio and optical catalogs were matched using the Astronomical Data Query Language at the \textit{Gaia} Archive web page\footnote{\url{https://gea.esac.esa.int/archive/}} with a search radius of 1 arcsecond. As a result, we found 12,700 sources with both optical and radio counterparts. Of these, only 107 sources had more than one optical source within 1 arcsecond of the radio position. In such cases, we chose the closest optical source as the match.

Studies aimed at improving the alignment between radio and optical reference systems often introduce additional sample filtering based on \textit{Gaia} astrometric solution quality flags \citep[e.g.,][]{Makarov2022,Lambert2024b}. To ensure the robustness of our results, we applied a filter selecting only sources with a renormalized unit weight error (RUWE) $ \le 1.4$. This cut only excluded nine sources from our final sample with polarization measurements, without qualitatively affecting the analysis results. However, such filtering may not only remove spurious sources with large VGD (e.g., where a foreground star is projected close to the AGN), it may exclude sources with genuine optical AGN substructures resolved by Gaia. Since the latter are the focus of our study, we opted to follow the approach of other works studying the phenomenon of VGD itself \citep[e.g.,][]{Kovalev2017, Xu2021} and did not apply any additional filtering.

   \begin{figure}
   \centering
   \includegraphics[width=0.95\hsize]{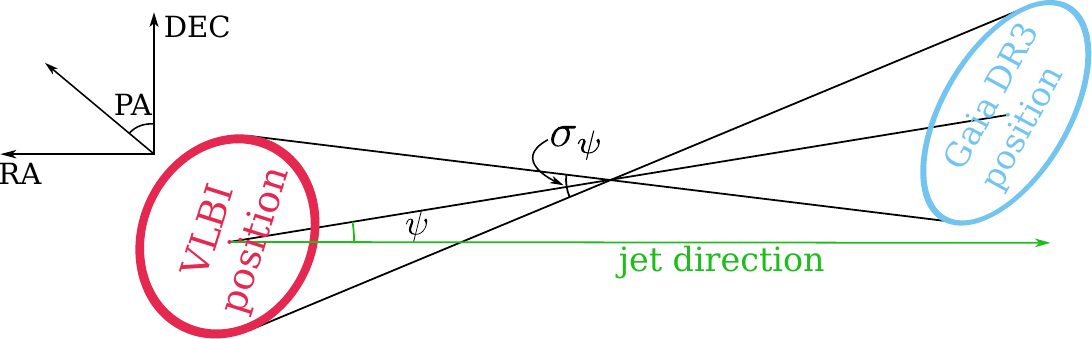}
      \caption{Sketch demonstrating quantities used in the analysis. The red and the blue ellipses denote $1\sigma$ positional uncertainty of a source in VLBI and \textit{Gaia} DR3 data, respectively. The green vector represents the parsec-scale jet direction. $\psi$ is the angle between the jet direction and the VGD vector. $\sigma_{\psi}$ is the estimate of uncertainty of $\psi$, calculated as the angle between maximum and minimum $\psi$ consistently with the error ellipses.}
         \label{fig:uncert}
   \end{figure}

In the next step, we selected objects with significant displacements between VLBI and \textit{Gaia} positions. We calculated $\psi$, the angle between the jet direction and the vector connecting the VLBI and \textit{Gaia} positions. To determine the uncertainty of this angle, we constructed tangents connecting the extreme points of the error ellipses for the VLBI and \textit{Gaia} positions, as shown in Fig.~\ref{fig:uncert}. We identified the line segments connecting points on the two error ellipses with the maximum and minimum position angles. Then, we computed the opening angle $\sigma_\psi$ between these extreme tangent lines and selected objects with a significant VGD by setting the threshold criterion to $\sigma_\psi \le 35\dg$, which resulted in 2,558 sources. We adopted this approach to define significant VGD, closely following previous works \citep{Kovalev2017, Plavin2019, Kovalev2020}, to ensure direct comparability of results. This approach is also advantageous for further analysis, as $\sigma_\psi$ not only characterizes the significance of the arc length between the VLBI and \textit{Gaia} positions but also reflects the uncertainty of its position angle. We confirmed that $\sigma_\psi$ is closely related to the normalized separations \citep[as defined by Eq. 4 in][]{Mignard2016} between optical and radio coordinates. For sources with $\sigma_\psi \le 179\dg$, the correlation coefficient between $\sigma_\psi$ and the normalized separation is -0.94 on a logarithmic scale. Therefore, by selecting sources with $\sigma_\psi \le 35\dg$, we effectively choose sources with the largest normalized separations.
Finally, we classified the selected sources with significant VGDs in three groups:
\begin{enumerate}
      \item $\psi=0\dg$ -- Downstream displacement sample containing objects where the VGD vector is co-aligned with the parsec-scale jet; that is, $\psi \in (-45\dg, +45\dg)$
      \item $\psi=180\dg$ -- Upstream displacement sample containing objects with the VGD and the jet pointing in the opposite directions; that is, $\psi \in (135\dg, 225\dg)$
      \item Objects with any other $\psi$
   \end{enumerate}

\section{Dependence of optical polarization in the VGD direction} \label{sec:pol_vs_vg}

\begin{figure*}
   \includegraphics[width=0.85\hsize]{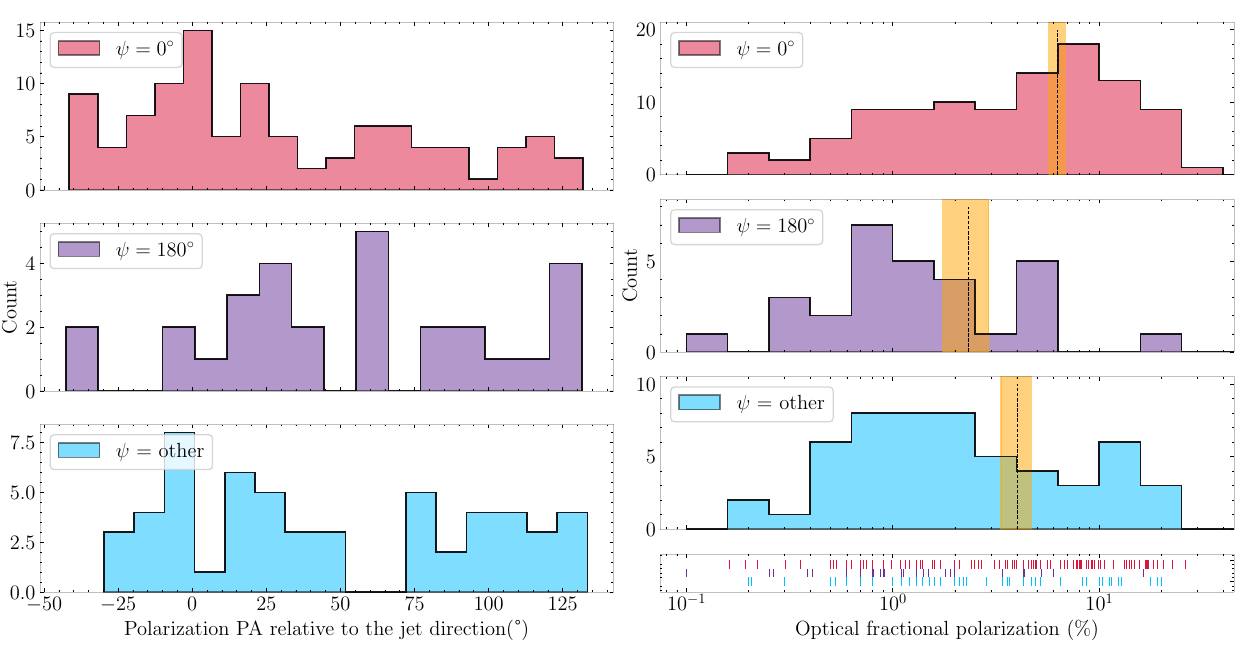}
      \caption{Distribution of difference between position angles of optical polarization plane and jet direction (left column) and fractional polarization (right column) for sources of three different classes according to their VGD vector direction. The optical polarimetric data used in this plot include our polarimetric data joined with the data used by \cite{Kovalev2020}. The vertical lines and orange regions indicate the sample mean and standard error. We note the logarithmic scale of the fractional polarization axis.}
         \label{fig:hist_subset}
\end{figure*}

We combined our polarimetric measurements described in Sects.~\ref{subsec:FORS}-\ref{subsec:CAHA} with the polarization data from \cite{Kovalev2020} and merged this catalog with the list of AGNs with significant VGDs. This resulted in a sample comprising 187 sources. We then constructed distributions of the optical fractional polarization and the polarization plane position angle with respect to the jet direction for the three samples: downstream VGD ($\psi=0\dg$), upstream VGD ($\psi=180\dg$), and sources with all other possible values of $\psi$. These distributions are shown in Fig.~\ref{fig:hist_subset}.
The sample's average fractional polarization was found to be $6.3\pm0.6$\%, $2.3\pm0.6$\%, and $4.0\pm0.7$\% for $\psi=0\dg$, $\psi=180\dg,$ and other $\psi$, respectively. According to the Student's t test, the average fractional polarization of the downstream and upstream samples can be equal with the p value = 0.0009 ($3.3\sigma$). Additionally, we performed the two-sample Kolmogorov-Smirnov (KS) test to check whether the fractional polarization distributions of the $\psi=0\dg$ and $\psi=180\dg$ samples could originate from the same underlying probability distribution. For this test, we obtained the p value = 0.0004 ($3.5\sigma$). Thus, both tests demonstrate that AGNs with significant VGDs pointing down the parsec-scale jet on average exhibit significantly higher optical polarization compared to the opposite case, where the VGD is pointing toward the central engine.

The distribution of the difference between the position angles of the polarization plane and the jet direction for the downstream sample peaks at $0\dg$. To verify the nonuniformity of this distribution, we applied Kuiper's test \citep{Kuiper1960} with the null hypothesis that the orientation of the polarization plane with respect to the jet direction is uniform. The test returned a p value=0.002, which rejects the null hypothesis at a $3.1\sigma$ significance level.

  \begin{figure*}
     \includegraphics[width=0.85\hsize]{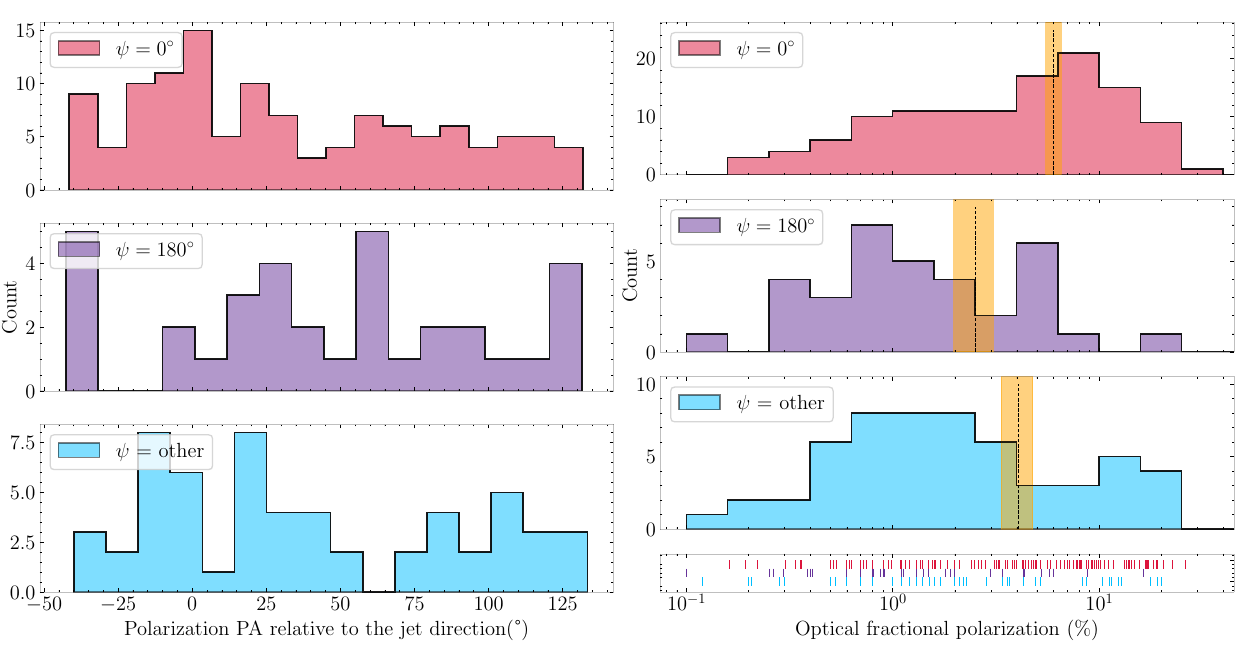}
        \caption{Same as Fig.~\ref{fig:hist_subset}, but for the entire polarimetric sample comprising our measurements combined with all available archival AGN polarization measurements from \cite{Kovalev2020} and \cite{Friedman2020}.}
        \label{fig:hist_all}
  \end{figure*}

We repeated the same analysis for the larger, but likely noisier, entire polarimetric sample (see Sect.~\ref{subsec:arch}). By merging this dataset with the list of AGNs exhibiting significant VGD, we obtained a sample of 211 sources. The distributions of polarization parameters for the three subsamples $\psi=0\dg$, $\psi=180\dg,$ and other $\psi$ in this dataset are shown in Fig.~\ref{fig:hist_all}. Qualitatively, they demonstrate the same dependencies as the "cleaner" sample discussed earlier. The mean fractional polarization of the downstream sample appears to be higher than in the sample with upstream VGD, while the distribution of the polarization angle relatively to the jet direction also shows a possible peak near $0\dg$.

We performed the same set of statistical tests to assess the significance of these features. The mean fractional polarization for the three samples $\psi=0\dg$, $\psi=180\dg,$ and other $\psi$ is $6.0\pm0.5$\%, $2.5\pm0.6$\%, and $4.1\pm0.7$\%. The Student's t-test provided a p value of 0.0009 ($3.3\sigma$), indicating that the probability of the mean fractional polarization being equal in the upstream and downstream samples is low. Additionally, the KS test indicated that the distributions of the fractional polarization in the two samples are likely inconsistent, with a p value of 0.0017 ($3.1\sigma$). However, the peak in the angle between the polarization plane and the jet direction for the downstream sample is not as significant as in the "clean" polarization sample. The results of the Kuiper test, with a p value of 0.009 ($2.6\sigma$), demonstrate that the distribution of the relative polarization PA does not deviate significantly from a uniform distribution.

Overall, we conclude that the sample of AGNs with significant VGDs pointing downstream of the jet has, on average, higher optical polarization compared to the upstream sample. There is a slight tendency for the downstream sources to align their optical polarization plane with the jet direction.

\section{Dependence of millimeter-wavelength polarization in the VGD direction} \label{sec:mmp_vs_vg}

\begin{figure}
     \includegraphics[width=0.95\hsize]{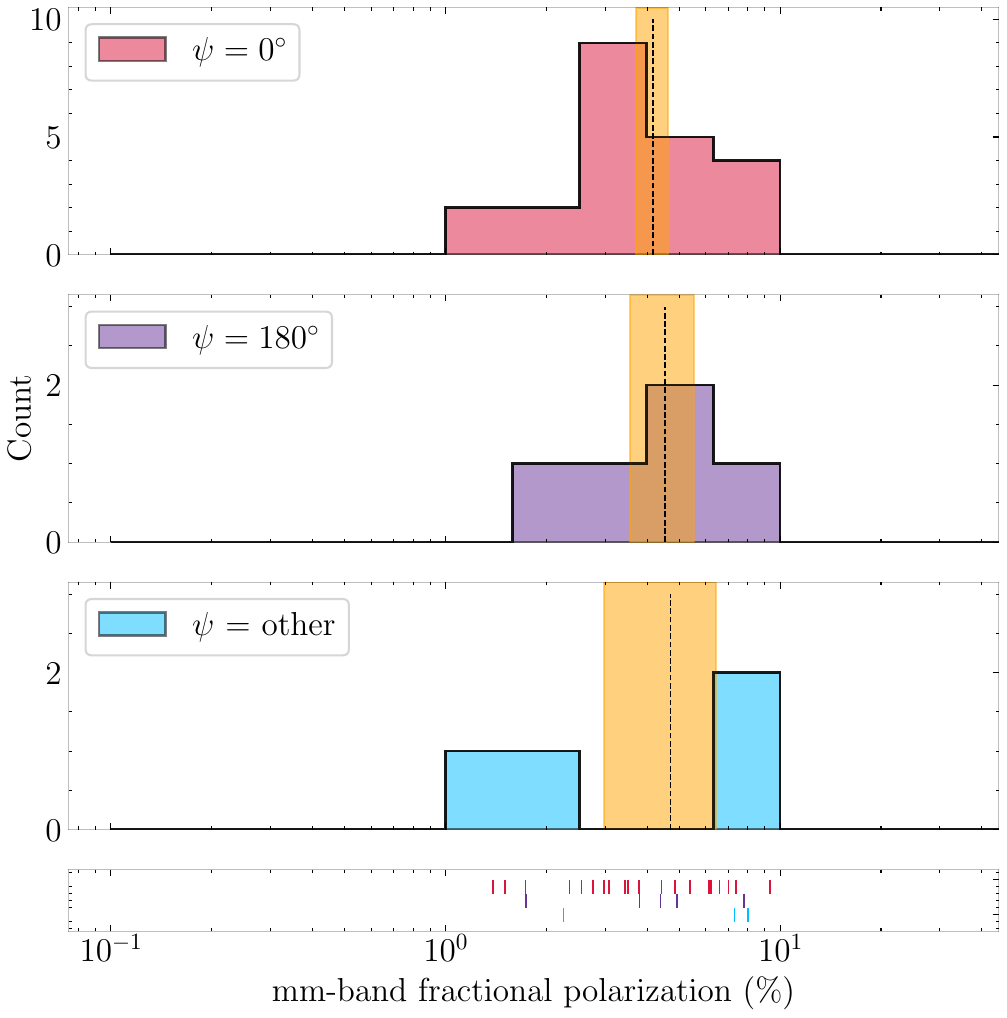}
        \caption{Distribution of 44 GHz fractional polarization among sources of the three samples. The vertical lines and orange regions indicate the sample mean and its standard error.}
        \label{fig:hist_pd_mm}
  \end{figure}

We cross-matched the catalog of polarization measurements at 44 GHz by the \textit{Planck} observatory (see Sect.~\ref{subsec:arch}) with the catalog of AGNs with significant VGDs. This provided 31 common sources. The distribution of the fractional polarization for the three subsamples classified based on their VGD vector direction are shown in Fig.~\ref{fig:hist_pd_mm}.

Contrary to the optical band, the millimeter-wavelength polarization does not tend to be higher for sources with downstream VGD compared to other samples. The mean values of the polarization are $4.2\pm0.5$\% and $4.5\pm1.0$\% of the $\psi=0\dg$ and $\psi=180\dg$. These values are consistent within uncertainties and are also consistent according to the Student's t-test (p value = 0.7). Moreover, according to the KS test these distributions of millimeter-wavelength polarization for the downstream and upstream samples can originate from the same parent distribution (p value = 0.5).
Thereby, we conclude that the millimeter-band fractional polarization of AGNs with significant VGDs does not depend on the orientation of this displacement with respect to the parsec-scale jet direction. 

\section{Dependence of the optical polarization spectrum on the redshift} \label{sec:pol_vs_z}

\cite{Plavin2019} found that there are differences in distributions of the redshift and the optical color index, $\Delta m$, between AGN samples with downstream and upstream VGD. Sources with $\psi=0\dg$ are distributed rather uniformly by the redshift and $\Delta m \in [0.5,1.6]$, while the majority of AGNs with $\psi=180\dg$ have $z > 1.1$ and a relatively blue $\Delta m \lesssim 0.8^m$ spectrum. This finding suggests that upstream VGD occurs when the central engine, particularly the accretion disk emission, dominates total optical flux of the AGN. The accretion disk emits primarily through a thermal mechanism, resulting in a narrow peak known as the big blue bump (BBB), which typically peaks in the UV range of the spectrum \citep[$\sim 110$ nm, ][]{Shang2005} for $z = 0$. For low-redshift sources, this BBB emission does not enter \textit{Gaia}'s pass band \citep{Evans2018}. Therefore, such sources with dominant accretion disks ($\psi=180^\circ$) are absent from the sample of AGNs with significant VGDs. As the redshift increases, the BBB emission shifts toward the pass band of \textit{Gaia} and starts entering it after $z \gtrsim 1.1$. At this point, if the accretion disk is luminous enough, it can dominate the overall emission in the \textit{Gaia} pass band, causing the upstream displacement of the optical photocenter and producing a blue spectral index.

We attempted to identify this effect in polarization. For this purpose, we conducted polarimetric observations of sources distributed along the redshift using two optical filters: the SDSS-$g^\prime$ and the Cousins-R (see Sects.~\ref{subsec:RoboPol} and \ref{subsec:CAHA}). These filters roughly approximate the blue sides of \textit{Gaia}'s BP and RP bands \citep{Ritter2020}. Given the location of the blue edges of the $g^\prime$ and R bands, the BBB emission should enter these bands at $z \gtrsim 1.1$ and $z \gtrsim 2.6$, respectively. A few sources with upstream displacements and low redshifts present in the catalog are likely AGNs for which the inner jet located closer to the central engine than the radio core dominates in the optical. Therefore, for low redshifts ($z < 1.1$), the ratio of polarization measured in the two bands, $PD(g^\prime)/PD(R)$, is expected to be around unity for both upstream and downstream samples. This is because the optical emission in both cases is produced by the synchrotron mechanism in the jet, which has a weak dependence of polarization on the wavelength in the considered range \citep{Marscher2021}. For sources in the $ 1.1 \lesssim z \lesssim 2.6$ range, the BBB emission enters the $g^\prime$ band (as well as Gaia's pass band) but does not yet reach the R band. Consequently, a significant fraction of sources with $\psi=180\dg$ in this redshift range likely have an accretion disk dominating their total flux. Thus, the $PD(g^\prime)/PD(R)$ ratio should drop because the $g^\prime$ band polarization is produced by scattering the in the accretion disk; whereas in the R band, it is still of synchrotron origin.

\begin{figure}
   \centering
   \includegraphics[width=0.95\hsize]{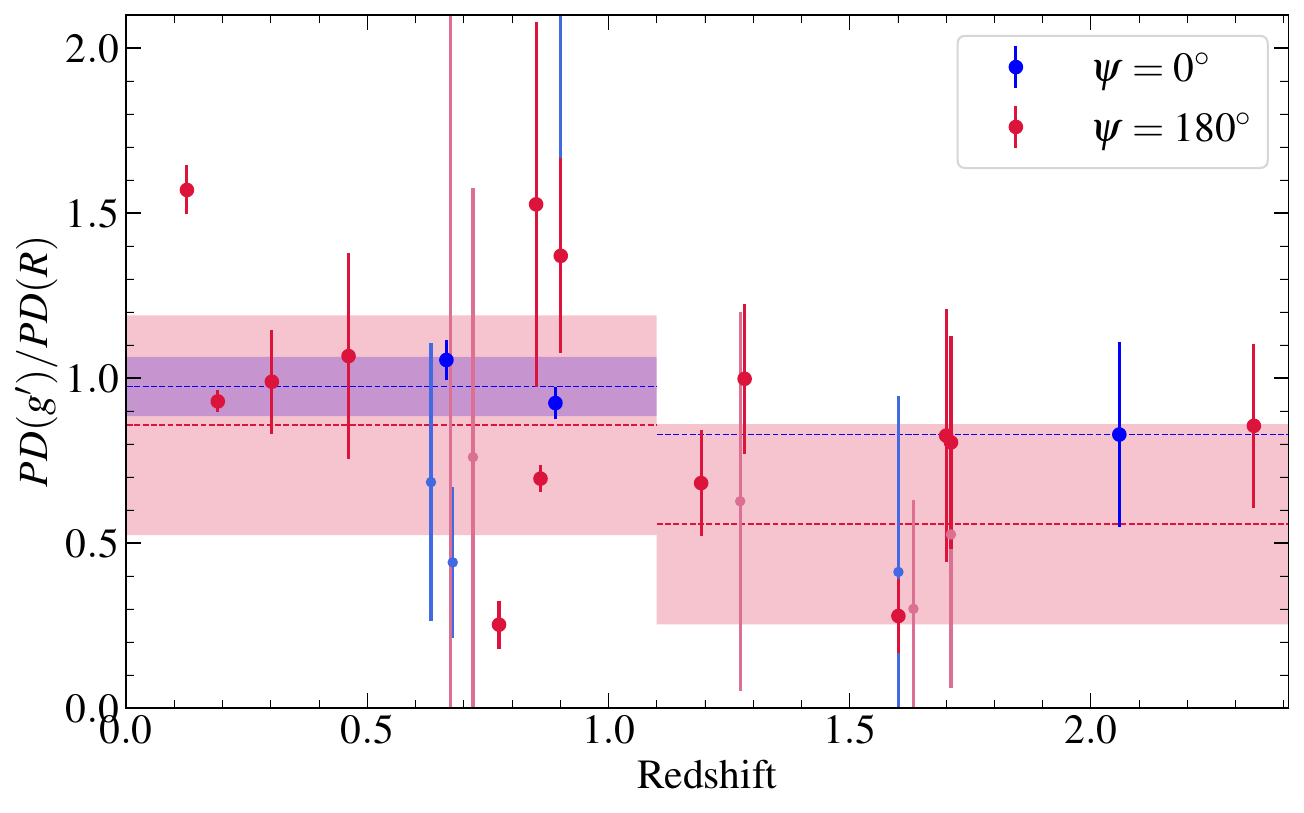}
      \caption{Dependence of ratio of fractional polarization in $g^\prime$ and R bands on redshift. The blue and the red dashed lines show the weighted mean values for corresponding samples in the ranges $0 < z < 1.1$ and $1.1 \le z < 2.4$. The filled regions indicate their uncertainties. The faded points are excluded from the averaging due to large uncertainties (S/N < 2).}
         \label{fig:wavedep}
   \end{figure}

The observed dependence of $PD(g^\prime)/PD(R)$ on the redshift is shown in Fig.~\ref{fig:wavedep}.
Although there is a slight decrease in the weighted average polarization for the upstream sample at $z \gtrsim 1.1$, the uncertainties are too large to draw a definitive conclusion. Therefore, we infer that no significant decrease in the ratio of fractional polarization between the two bands was detected for $z \gtrsim 1.1$ in sources with upstream VGD. More accurate observations of a larger sample are needed to verify the hypothesis that accretion-disk emission affects polarization within a certain redshift range.

\section{Discussion and conclusions} \label{sec:concl}

We analyzed our optical polarization measurements of 152 AGNs joined with archival polarimetric data and combined with data on VGDs and the parsec-scale jet position angles. We confirmed the previously reported dependence of optical polarization on the VGD position angle with respect to the jet direction \citep{Kovalev2020}. Given the larger size of the optical polarization catalog and the more precise \textit{Gaia} DR3 astrometry compared to the previous study, we conclude that AGNs with VGDs pointing down the jet on average exhibit significantly higher optical fractional polarization compared to sources with upstream displacements. This finding aligns well with the model previously proposed to explain VGD \citep{Kovalev2017,Plavin2019}. Displacements pointing down the jet are produced by the extended optical jets, whose optical photocenters are located downstream relatively to the radio core. In such cases, the optical emission is highly polarized due to its synchrotron nature. There is a slight tendency for these sources to align their polarization plane with the jet direction, which could indicate that the toroidal magnetic field component prevails in their jets \citep{Lyutikov2005}. In contrast, for upstream VGDs the accretion disk emission dominates the total optical flux, which results in lower polarization due to scattering in a configuration close to axial symmetry.

Furthermore, this model is supported by the analysis of millimeter-band polarization in Sect.~\ref{sec:mmp_vs_vg}. Contrary to optical emission, the 44 GHz polarization shows similar distributions in the downstream and upstream VGD samples. This is explained by the fact that in both samples we only observe the jet emission in this band, while the BBB produced by the accretion disk is located at much higher frequencies. Thus, regardless of whether an AGN exhibits a prominent accretion disk or a strong jet in the optical, in the millimeter band we receive the synchrotron radiation, which possesses the same polarization in both cases.

We propose an additional observational test in Sect.~\ref{sec:pol_vs_z}, which could provide further evidence for the considered model. The fractional polarization in the $g^\prime$ filter is expected to decrease relative to that in the R filter when accretion disk emission enters the former band pass at $z \gtrsim 1.1$. However, the small sample size and the relatively low signal-to-noise ratio of our two-band measurements did not allow us to detect the expected changes.

While the causes of upstream and downstream VGD appear to be understood, the reasons behind the dominance of jet or disk emission in some AGNs, but not in others, have yet to be revealed. Additionally, it remains unclear whether AGNs with insignificant VGDs possess distinct properties or represent an intermediate case between upstream and downstream VGD samples. To explore these issues further, we obtained the classifications and Doppler factors for our sample sources from the MILLIQUAS v.8 catalog \citep{Flesch2024} and \cite{Liodakis2018}. We found that our sample consists almost exclusively of type I AGNs. In the entire polarimetric sample, only three sources in the downstream VGD and one source in the upstream VGD samples are classified as type II AGNs. However, this is likely due to a selection bias. The average Doppler factors for the upstream ($8.3 \pm 1.5$ for 25 sources) and downstream ($8.8 \pm 1.4$ for 47 sources) VGD samples are consistent within uncertainties, while sources with insignificant VGDs have a substantially higher average Doppler factor of $14.3 \pm 1.2$ (109 sources). The average optical polarization for sources with insignificant VGD is $4.5\pm0.3$\%, which falls between the $2.5\pm0.6$\% and $6.0\pm0.5$\% found in Sect.~\ref{sec:pol_vs_vg} for the upstream and downstream VGD samples. These values suggest a preliminary interpretation consistent with recent findings by \cite{Lambert2024b}. AGNs with minimal or insignificant VGDs might have the smallest viewing angles between our line of sight and the jet among all AGNs. This could explain their high Doppler factors and the small projected angular distance between the radio core and the optical photocenter. Additionally, for small viewing angles ($\lesssim 6^\circ$), a smaller viewing angle should lead to a lower level of optical fractional polarization if the jet has a helical magnetic field \citep{Lyutikov2005}. This may explain why, on average, the optical polarization is lower in sources with insignificant VGDs compared to those with downstream VGDs, despite the higher dominance of jet emission due to Doppler boosting in these sources.

Finally, we speculate that aside from the time-averaged quantities used in this work, future \textit{Gaia} data releases with time-series positional data will offer further opportunities to resolve optical subcomponents in parsec-scale jets. As demonstrated by \cite{Blinov2021b}, individual emission features can dominate the entire AGN optical flux, while propagating many parsecs down the jet and causing the rotation of the optical polarization plane. Such events not only have the capacity to induce random jitter in the optical source position \citep{Petrov2019}, they are also expected to manifest as coherent changes of VGD over time in some AGNs. Therefore, optical polarization monitoring of sources with significant proper motions in \textit{Gaia} data is highly desirable.

\section{Data availability} \label{sec:data_av}

The new polarization measurements discussed in Sects.~\ref{subsec:FORS}-\ref{subsec:CAHA} are available in the Harvard Dataverse\footnote{\url{https://doi.org/10.7910/DVN/DJZAAV}}. Other data analyzed in this work can be accessed via the links and references provided in the text.

\begin{acknowledgements}
D.B. acknowledges support from the European Research Council (ERC) under the Horizon ERC Grants 2021 programme under the grant agreement No. 101040021. The work of A.A. was performed as part of the SAO RAS government contract approved by the Ministry of Science and Higher Education of the Russian Federation. Based on observations collected at the Centro Astronómico Hispánico en Andalucía (CAHA), operated jointly by the Instituto de Astrofisica de Andalucia (CSIC) and the Andalusian Universities (Junta de Andalucía).
\end{acknowledgements}

\bibliographystyle{aa}
\bibliography{bibliography.bib}

\begin{thebibliography}{42}
\expandafter\ifx\csname natexlab\endcsname\relax\def\natexlab#1{#1}\fi

\bibitem[{{Agol}(1997)}]{Agol1997}
{Agol}, E. 1997, PhD thesis, University of California, Santa Barbara

\bibitem[{{Blinov} {et~al.}(2020){Blinov}, {Casadio}, {Mandarakas}, \&
  {Angelakis}}]{Blinov2020}
{Blinov}, D., {Casadio}, C., {Mandarakas}, N., \& {Angelakis}, E. 2020, \aap,
  635, A102

\bibitem[{{Blinov} {et~al.}(2021{\natexlab{a}}){Blinov}, {Jorstad}, {Larionov},
  {MacDonald}, {Grishina}, {Kopatskaya}, {Larionova}, {Larionova}, {Morozova},
  {Nikiforova}, {Savchenko}, {Troitskaya}, \& {Troitsky}}]{Blinov2021b}
{Blinov}, D., {Jorstad}, S.~G., {Larionov}, V.~M., {et~al.} 2021{\natexlab{a}},
  \mnras, 505, 4616

\bibitem[{{Blinov} {et~al.}(2021{\natexlab{b}}){Blinov}, {Kiehlmann},
  {Pavlidou}, {Panopoulou}, {Skalidis}, {Angelakis}, {Casadio}, {Einoder},
  {Hovatta}, {Kokolakis}, {Kougentakis}, {Kus}, {Kylafis}, {Kyritsis},
  {Lalakos}, {Liodakis}, {Maharana}, {Makrydopoulou}, {Mandarakas},
  {Maragkakis}, {Myserlis}, {Papadakis}, {Paterakis}, {Pearson}, {Ramaprakash},
  {Readhead}, {Reig}, {S{\l}owikowska}, {Tassis}, {Xexakis}, {{\.Z}ejmo}, \&
  {Zensus}}]{Blinov2021a}
{Blinov}, D., {Kiehlmann}, S., {Pavlidou}, V., {et~al.} 2021{\natexlab{b}},
  \mnras, 501, 3715

\bibitem[{{Blinov} {et~al.}(2023){Blinov}, {Maharana}, {Bouzelou}, {Casadio},
  {Gjerl{\o}w}, {Jormanainen}, {Kiehlmann}, {Kypriotakis}, {Liodakis},
  {Mandarakas}, {Markopoulioti}, {Panopoulou}, {et~al.}}]{Blinov2023}
{Blinov}, D., {Maharana}, S., {Bouzelou}, F., {et~al.} 2023, \aap, 677, A144

\bibitem[{{Charlot} {et~al.}(2020){Charlot}, {Jacobs}, {Gordon}, {Lambert}, {de
  Witt}, {B{\"o}hm}, {Fey}, {Heinkelmann}, {Skurikhina}, {Titov}, {Arias},
  {Bolotin}, {Bourda}, {Ma}, {Malkin}, {Nothnagel}, {Mayer}, {MacMillan},
  {Nilsson}, \& {Gaume}}]{Charlot2020}
{Charlot}, P., {Jacobs}, C.~S., {Gordon}, D., {et~al.} 2020, \aap, 644, A159

\bibitem[{{Evans} {et~al.}(2018){Evans}, {Riello}, {De Angeli}, {Carrasco},
  {Montegriffo}, {Fabricius}, {Jordi}, {Palaversa}, {Diener}, {Busso},
  {Cacciari}, {van Leeuwen}, {Burgess}, {Davidson}, {Harrison}, {Hodgkin},
  {Pancino}, {Richards}, {Altavilla}, {Balaguer-N{\'u}{\~n}ez}, {Barstow},
  {Bellazzini}, {Brown}, {Castellani}, {Cocozza}, {De Luise}, {Delgado},
  {Ducourant}, {Galleti}, {Gilmore}, {Giuffrida}, {Holl}, {Kewley}, {Koposov},
  {Marinoni}, {Marrese}, {Osborne}, {Piersimoni}, {Portell}, {Pulone},
  {Ragaini}, {Sanna}, {Terrett}, {Walton}, {Wevers}, \&
  {Wyrzykowski}}]{Evans2018}
{Evans}, D.~W., {Riello}, M., {De Angeli}, F., {et~al.} 2018, \aap, 616, A4

\bibitem[{{Flesch}(2024)}]{Flesch2024}
{Flesch}, E.~W. 2024, The Open Journal of Astrophysics, 7, 6

\bibitem[{{Friedman} {et~al.}(2020){Friedman}, {Gerasimov}, {Leon}, {Stevens},
  {Tytler}, {Keating}, \& {Kislat}}]{Friedman2020}
{Friedman}, A.~S., {Gerasimov}, R., {Leon}, D., {et~al.} 2020, \prd, 102,
  043008

\bibitem[{{Gaia Collaboration} {et~al.}(2022){Gaia Collaboration}, {Klioner},
  {Lindegren}, {Mignard}, {Hern{\'a}ndez}, {Ramos-Lerate}, {Bastian},
  {Biermann}, {Bombrun}, {de Torres}, {Gerlach}, {Geyer}, {Hilger}, {Hobbs},
  {Lammers}, {et~al.}}]{Klioner2022}
{Gaia Collaboration}, {Klioner}, S.~A., {Lindegren}, L., {et~al.} 2022, \aap,
  667, A148

\bibitem[{{Gaia Collaboration} {et~al.}(2016){Gaia Collaboration}, {Prusti},
  {de Bruijne}, {Brown}, {Vallenari}, {Babusiaux}, {Bailer-Jones}, {Bastian},
  {Biermann}, {Evans}, {Eyer}, {Jansen}, {Jordi}, {Klioner},
  {et~al.}}]{Gaia2016}
{Gaia Collaboration}, {Prusti}, T., {de Bruijne}, J.~H.~J., {et~al.} 2016,
  \aap, 595, A1

\bibitem[{{Gaia Collaboration} {et~al.}(2023){Gaia Collaboration}, {Vallenari},
  {Brown}, {Prusti}, {de Bruijne}, {Arenou}, {Babusiaux}, {Biermann},
  {Creevey}, {Ducourant}, {Evans}, {Eyer}, {et~al.}}]{GaiaDR3}
{Gaia Collaboration}, {Vallenari}, A., {Brown}, A.~G.~A., {et~al.} 2023, \aap,
  674, A1

\bibitem[{{Ginzburg}(1979)}]{Ginzburg1979}
{Ginzburg}, V.~L. 1979, {Theoretical physics and astrophysics}, International
  Series on Nuclear Energy (Oxford: Pergamon)

\bibitem[{{Hutsem{\'e}kers} {et~al.}(2018){Hutsem{\'e}kers}, {Borguet},
  {Sluse}, \& {Pelgrims}}]{Hutsemekers2018}
{Hutsem{\'e}kers}, D., {Borguet}, B., {Sluse}, D., \& {Pelgrims}, V. 2018,
  \aap, 620, A68

\bibitem[{{Hutsem{\'e}kers} {et~al.}(2005){Hutsem{\'e}kers}, {Cabanac}, {Lamy},
  \& {Sluse}}]{Hutsemekers2005}
{Hutsem{\'e}kers}, D., {Cabanac}, R., {Lamy}, H., \& {Sluse}, D. 2005, \aap,
  441, 915

\bibitem[{{Itoh} {et~al.}(2016){Itoh}, {Nalewajko}, {Fukazawa}, {Uemura},
  {Tanaka}, {Kawabata}, {Madejski}, {Schinzel}, {Kanda}, {Shiki}, {Akitaya},
  {Kawabata}, {Moritani}, {Nakaoka}, {Ohsugi}, {Sasada}, {Takaki}, {Takata},
  {Ui}, {Yamanaka}, \& {Yoshida}}]{Itoh2016}
{Itoh}, R., {Nalewajko}, K., {Fukazawa}, Y., {et~al.} 2016, \apj, 833, 77

\bibitem[{{King} {et~al.}(2014){King}, {Blinov}, {Ramaprakash}, {Myserlis},
  {Angelakis}, {Balokovi{\'c}}, {Feiler}, {Fuhrmann}, {Hovatta}, {Khodade},
  {Kougentakis}, {Kylafis}, {Kus}, {Modi}, {Paleologou}, {Panopoulou},
  {Papadakis}, {Papamastorakis}, {Paterakis}, {Pavlidou}, {Pazderska},
  {Pazderski}, {Pearson}, {Rajarshi}, {Readhead}, {Reig}, {Steiakaki},
  {Tassis}, \& {Zensus}}]{King2014}
{King}, O.~G., {Blinov}, D., {Ramaprakash}, A.~N., {et~al.} 2014, \mnras, 442,
  1706

\bibitem[{{Kovalev} {et~al.}(2017){Kovalev}, {Petrov}, \&
  {Plavin}}]{Kovalev2017}
{Kovalev}, Y.~Y., {Petrov}, L., \& {Plavin}, A.~V. 2017, \aap, 598, L1

\bibitem[{{Kovalev} {et~al.}(2020){Kovalev}, {Zobnina}, {Plavin}, \&
  {Blinov}}]{Kovalev2020}
{Kovalev}, Y.~Y., {Zobnina}, D.~I., {Plavin}, A.~V., \& {Blinov}, D. 2020,
  \mnras, 493, L54

\bibitem[{{Kuiper}(1960)}]{Kuiper1960}
{Kuiper}, N.~H. 1960, Proceedings of the Koninklijke Nederlandse Akademie van
  Wetenschappen, Series A, 63, 38

\bibitem[{{Lambert} \& {Secrest}(2024)}]{Lambert2024b}
{Lambert}, S. \& {Secrest}, N.~J. 2024, \aap, 684, A93

\bibitem[{{Lambert} {et~al.}(2024){Lambert}, {Sol}, \& {Pierron}}]{Lambert2024}
{Lambert}, S., {Sol}, H., \& {Pierron}, A. 2024, \aap, 684, A202

\bibitem[{{Liodakis} {et~al.}(2018){Liodakis}, {Hovatta}, {Huppenkothen},
  {Kiehlmann}, {Max-Moerbeck}, \& {Readhead}}]{Liodakis2018}
{Liodakis}, I., {Hovatta}, T., {Huppenkothen}, D., {et~al.} 2018, \apj, 866,
  137

\bibitem[{{Lyutikov} {et~al.}(2005){Lyutikov}, {Pariev}, \&
  {Gabuzda}}]{Lyutikov2005}
{Lyutikov}, M., {Pariev}, V.~I., \& {Gabuzda}, D.~C. 2005, \mnras, 360, 869

\bibitem[{{Makarov} {et~al.}(2019){Makarov}, {Berghea}, {Frouard}, {Fey}, \&
  {Schmitt}}]{Makarov2019}
{Makarov}, V.~V., {Berghea}, C.~T., {Frouard}, J., {Fey}, A., \& {Schmitt},
  H.~R. 2019, \apj, 873, 132

\bibitem[{{Makarov} \& {Secrest}(2022)}]{Makarov2022}
{Makarov}, V.~V. \& {Secrest}, N.~J. 2022, \apj, 933, 28

\bibitem[{{Malkin}(2018)}]{Malkin2018}
{Malkin}, Z. 2018, \apjs, 239, 20

\bibitem[{{Mandarakas} {et~al.}(2021){Mandarakas}, {Blinov}, {Casadio},
  {Pelgrims}, {Kiehlmann}, {Pavlidou}, \& {Tassis}}]{Mandarakas2021}
{Mandarakas}, N., {Blinov}, D., {Casadio}, C., {et~al.} 2021, \aap, 653, A123

\bibitem[{{Marscher} \& {Jorstad}(2021)}]{Marscher2021}
{Marscher}, A.~P. \& {Jorstad}, S.~G. 2021, Galaxies, 9, 27

\bibitem[{{Mignard} {et~al.}(2016){Mignard}, {Klioner}, {Lindegren}, {Bastian},
  {Bombrun}, {Hern{\'a}ndez}, {Hobbs}, {Lammers}, {Michalik}, {Ramos-Lerate},
  {Biermann}, {Butkevich}, {Comoretto}, {Joliet}, {Holl}, {Hutton}, {Parsons},
  {Steidelm{\"u}ller}, {Andrei}, {Bourda}, \& {Charlot}}]{Mignard2016}
{Mignard}, F., {Klioner}, S., {Lindegren}, L., {et~al.} 2016, \aap, 595, A5

\bibitem[{{Patat} \& {Romaniello}(2006)}]{Patat2006}
{Patat}, F. \& {Romaniello}, M. 2006, \pasp, 118, 146

\bibitem[{{Petrov} \& {Kovalev}(2017{\natexlab{a}})}]{Petrov2017a}
{Petrov}, L. \& {Kovalev}, Y.~Y. 2017{\natexlab{a}}, \mnras, 471, 3775

\bibitem[{{Petrov} \& {Kovalev}(2017{\natexlab{b}})}]{Petrov2017b}
{Petrov}, L. \& {Kovalev}, Y.~Y. 2017{\natexlab{b}}, \mnras, 467, L71

\bibitem[{{Petrov} {et~al.}(2019){Petrov}, {Kovalev}, \& {Plavin}}]{Petrov2019}
{Petrov}, L., {Kovalev}, Y.~Y., \& {Plavin}, A.~V. 2019, \mnras, 482, 3023

\bibitem[{{Plavin} {et~al.}(2019){Plavin}, {Kovalev}, \& {Petrov}}]{Plavin2019}
{Plavin}, A.~V., {Kovalev}, Y.~Y., \& {Petrov}, L.~Y. 2019, \apj, 871, 143

\bibitem[{{Plavin} {et~al.}(2022){Plavin}, {Kovalev}, \&
  {Pushkarev}}]{Plavin2022}
{Plavin}, A.~V., {Kovalev}, Y.~Y., \& {Pushkarev}, A.~B. 2022, \apjs, 260, 4

\bibitem[{{Ritter} \& {Huang}(2020)}]{Ritter2020}
{Ritter}, A. \& {Huang}, C. 2020, in Journal of Physics Conference Series, Vol.
  1593, Journal of Physics Conference Series (IOP), 012039

\bibitem[{{Rocha} {et~al.}(2023){Rocha}, {Keskitalo}, {Partridge}, {Marscher},
  {O'Dea}, {Pearson}, \& {G{\'o}rski}}]{Rocha2023}
{Rocha}, G., {Keskitalo}, R., {Partridge}, B., {et~al.} 2023, \aap, 669, A92

\bibitem[{{Secrest}(2022)}]{Secrest2022}
{Secrest}, N.~J. 2022, \apjl, 939, L32

\bibitem[{{Shang} {et~al.}(2005){Shang}, {Brotherton}, {Green}, {Kriss},
  {Scott}, {Quijano}, {Blaes}, {Hubeny}, {Hutchings}, {Kaiser}, {Koratkar},
  {Oegerle}, \& {Zheng}}]{Shang2005}
{Shang}, Z., {Brotherton}, M.~S., {Green}, R.~F., {et~al.} 2005, \apj, 619, 41

\bibitem[{{Whiting}(2000)}]{Whiting2000}
{Whiting}, M.~T. 2000, PhD thesis, University of Melbourne, Australia

\bibitem[{{Xu} {et~al.}(2021){Xu}, {Lunz}, {Anderson}, {Savolainen}, {Zubko},
  \& {Schuh}}]{Xu2021}
{Xu}, M.~H., {Lunz}, S., {Anderson}, J.~M., {et~al.} 2021, \aap, 647, A189

\end{thebibliography}

\end{document}